**RESEARCH**

**Open Access**

# Improvement of variables interpretability in kernel PCA

Mitja Briscik[1*], Marie-Agnès Dillies[2] and Sébastien Déjean[1]

*Correspondence:
mitja.briscik@math.univ-toulouse.fr

[1] Institut de Mathématiques de Toulouse, UMR5219, CNRS, UPS, Université de Toulouse, Cedex 9, 31062 Toulouse, France
[2] Institut Pasteur, Université Paris Cité, Bioinformatics and Biostatistics Hub, F-75015 Paris, France

## Abstract

**Background:** Kernel methods have been proven to be a powerful tool for the integration and analysis of high-throughput technologies generated data. Kernels offer a nonlinear version of any linear algorithm solely based on dot products. The kernelized version of principal component analysis is a valid nonlinear alternative to tackle the nonlinearity of biological sample spaces. This paper proposes a novel methodology to obtain a data-driven feature importance based on the *kernel PCA* representation of the data.

**Results:** The proposed method, kernel PCA Interpretable Gradient (KPCA-IG), provides a data-driven feature importance that is computationally fast and based solely on linear algebra calculations. It has been compared with existing methods on three benchmark datasets. The accuracy obtained using KPCA-IG selected features is equal to or greater than the other methods' average. Also, the computational complexity required demonstrates the high efficiency of the method. An exhaustive literature search has been conducted on the selected genes from a publicly available Hepatocellular carcinoma dataset to validate the retained features from a biological point of view. The results once again remark on the appropriateness of the computed ranking.

**Conclusions:** The black-box nature of kernel PCA needs new methods to interpret the original features. Our proposed methodology KPCA-IG proved to be a valid alternative to select influential variables in high-dimensional high-throughput datasets, potentially unravelling new biological and medical biomarkers.

## Background

The recent advancement in high-throughput biotechnologies is making large multi-omics datasets easily available. Bioinformatics has recently entered the *Big Data* era, offering researchers new perspectives to analyse biological systems to discover new genotype-phenotype interactions.

Consequently, new ad-hoc methods to optimise post-genomic data analysis are needed, considering the high complexity and heterogeneity involved. For instance, multi-omics datasets pose the additional difficulty of dealing with a multilayered framework making data integration extremely challenging.

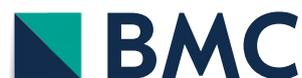

In this context, kernel methods offer a natural theoretical framework for the high dimensionality and heterogeneous nature of omics data, addressing their peculiar convoluted nature [63]. These methods facilitate the analysis and the integration of various types of omics data, such as vectors, sequences, networks, phylogenetic trees, and images, through a relevant kernel function. Using kernels enables the representation of the datasets in terms of pairwise similarities between sample points, which is helpful for handling high-dimensional sample spaces more efficiently than using Euclidean distance alone. Euclidean distance can be inadequate in complex scenarios, as stated in [17], but kernels can help overcome this limitation. Moreover, kernel methods have the advantage of providing a nonlinear version of any linear algorithm which relies solely on dot products. For instance, Kernel Principal Component Analysis [62], Kernel Canonical Correlation Analysis [4], Kernel Discriminant Analysis [53] and Kernel Clustering [21] are all examples of nonlinear algorithms enabled by the so-called kernel trick.

This work will focus on the kernelized version of Principal Component Analysis, KPCA, that provides a nonlinear alternative to the standard PCA to reduce the sample space dimensions.

However, KPCA and kernel methods, in general, pose new challenges in interpretability. The so-called *pre-image* problem arises since data points are only addressed through the kernel function, causing the original features to be lost during the data embedding process. The initial information contained in the original variables is summarised in the pairwise kernel similarity scores among data sample points. Thus, retrieving the original input dimensions is highly challenging when it comes to identifying the most prominent features. Even if it is possible for certain specific kernels to solve the pre-image problem through a fixed-point iteration method, the provided solution is typically numerically unstable since it involves a non-convex optimisation problem [59]. Moreover, in most cases, the exact pre-image does not even exist [43].

However, it is possible to find works that aim at finding the pre-image problem solution, like the pre-image based on distance constraints in the feature space in [32] or local isomorphism as in [25].

Instead, in this article, we propose KPCA Interpretable Gradient (*KPCA-IG*), a novel methodology to assess the contribution of the original variables on the KPCA solution, which is based on the computations of the partial derivatives of the kernel itself. More specifically, this method aims at identifying the most influential variables for the kernel principal components that account for the majority of the variability of the data. To the best of our knowledge KPCA-IG is the first method for KPCA that offers a computationally fast and stable data-driven feature ranking to identify the most prominent original variables, solely based on the norm of gradients computations. Consequently, unimportant descriptors can be ignored, refining the kernel PCA procedure whose similarity measure can be influenced by irrelevant dimensions [7].

**Existing approaches to facilitate feature interpretability in the unsupervised setting**

The literature on unsupervised feature selection is generally less extensive than its supervised learning counterpart. One of the main reasons for this disparity is that the selection is made without a specific prediction goal, making it difficult to evaluate the quality of a particular solution. In the same way, the unsupervised feature selection field that takes advantage of the kernel framework has been found to be less explored than kernel applications with classification purposes. As mentioned earlier, interpreting kernel PCA requires additional attention as the kernel principal component axes themselves are only defined by the similarity scores of the sample points. However, the literature has limited attempts to explain how to interpret these axes after the kernel transformation. Therefore, feature selection methods based on KPCA are rare.

Among others, [52] proposed a method to visualize the original variables into the 2*D* kernel PCs plot. For every sample point projected in the KPCA axes, they propose to display the original variables as arrows representing the vector field of the direction of maximum growth for each input variable or combination of them. This algorithm does not provide variable importance ranking, requiring previous knowledge about which variables to display.

On the contrary, [40] introduced a variable importance selection method to identify the most influential variables for every principal component based on random permutation. The procedure is performed for all variables, selecting the ones that result in the largest Crone-Crosby [14] distance between kernel matrices, i.e. the variables whose permutations of the observations lead to a significant change in the kernel Gram matrix values. However, the method does not come with a variable representation and can be computationally expensive, as many other permutation-based methods. This method will be denoted as *KPCA-permute* in the rest of the article.

Another method that takes advantage of the kernel framework is the unsupervised method *UKFS* with its extension *UKFS-KPCA* in [7] where the authors proposed to select important features through a non-convex optimization problem with a $\ell_1$ penalty for a Frobenius norm distortion measure.

As exhaustively described in the overview presented in [35], there are different approaches to assess variable importance in an unsupervised setting not based on the kernel framework. Among others, we can mention two methodologies that are based on the computation of a score, the Laplacian Score *lapl* in [23] and its extension Spectral Feature Selection *SPEC* in [78]. Other alternatives are the Multi-Cluster Feature Selection *MCFS* in [8], the Nonnegative Discriminative Feature Selection *NDFS* in [37], and the Unsupervised Discriminative Feature Selection *UDFS* in [75]. These methods aim to select features by keeping only the ones that best represent the implicit nature of the clustered data. Then, Convex Principal Feature Selection *CPFS* [41] adopts a distinct approach to feature selection, focusing on selecting a subset of features that can best reconstruct the projection of the data on the initial axes of the Principal Component Analysis.

As mentioned, the present study introduces a novel contribution to the interpretability of variables in kernel PCA, assuming that the first kernel PC axes contain the most relevant information about the data. The newly proposed method follows and extends the idea proposed by [52], with the fundamental difference that it gives a data-driven features importance ranking. Moreover, contrarily to *KPCA-permute* in [40], it does not have a random nature while being considerably faster.

## Methods

This section presents the formulation behind our proposed method *KPCA-IG*, starting with the description of the kernel framework.

### Kernel PCA

Given a dataset of $n$ observations $\boldsymbol{x}_1, \ldots, \boldsymbol{x}_n$ with $\boldsymbol{x}_i \in \chi$, a function k defined as k: $\chi \times \chi \longrightarrow \mathbb{R}$ is a valid kernel if it is symmetric and positive semi-definite i.e. $k(\boldsymbol{x}_i, \boldsymbol{x}_j) = k(\boldsymbol{x}_j, \boldsymbol{x}_i)$ and $\boldsymbol{c}^T \boldsymbol{K} \boldsymbol{c} \geqslant 0$, $\forall \boldsymbol{c} \in \mathbb{R}^n$, where $\boldsymbol{K}$ is the $n \times n$ kernel matrix containing all the data pairwise similarities $\boldsymbol{K} = k(x_i, x_j)$. The input set $\chi$ does not require any assumption. In this work we consider it to be $\chi = \mathbb{R}^p$.

Every kernel function is associated with an implicit function $\phi : \chi \longrightarrow \mathcal{H}$ which maps the input points into a generic feature space $\mathcal{H}$, with possibly an infinite dimensionality, with the expression $k(\boldsymbol{x}_i, \boldsymbol{x}_j) = \langle \phi(\boldsymbol{x}_i), \phi(\boldsymbol{x}_j) \rangle$. This relation allows to compute the dot products in the feature space, implicitly applying the kernel function to the input objects, without explicitly computing the mapping function $\phi$.

Principal Component Analysis is a well-established linear algorithm to extract the data structure in an unsupervised setting [22]. However, it is commonly accepted that in specific fields, such as bioinformatics, assuming a linear sample space may not help to capture the data manifold adequately [52]. In other words, the relationships between the variables may be nonlinear, making linear methods unsuitable. Hence, with high-dimensional data such as genomic data, where the number of features is usually much larger than the number of samples, nonlinear methods like kernel methods can provide a valid alternative for data analysis.

A compelling approach to overcome this challenge is through kernel PCA, which was introduced in [62]. Kernel PCA applies PCA in the feature space generated by the kernel, and as PCA relies on solving an eigenvalue problem, its kernelized version operates under the same principle.

The algorithm requires the data to be centered in the feature space, and the diagonalization of the centered covariance matrix in the feature space $\mathcal{H}$ is equivalent to the eigendecomposition of the kernel matrix $\boldsymbol{K}$. The data coordinates in the feature space are unknown as $\phi$ is not explicitly computed. Consequently, the required centering of variables in the feature space cannot be done directly. However, it is possible to compute the centered Gram matrix $\tilde{\boldsymbol{K}}$ as $\tilde{\boldsymbol{K}} = \boldsymbol{K} - \frac{1}{n} \boldsymbol{K} \boldsymbol{1}_n \boldsymbol{1}_n^T - \frac{1}{n} \boldsymbol{1}_n \boldsymbol{1}_n^T \boldsymbol{K} + \frac{1}{n^2} (\boldsymbol{1}_n^T \boldsymbol{K} \boldsymbol{1}_n) \boldsymbol{1}_n \boldsymbol{1}_n^T$ with $\boldsymbol{1}_n$ a vector with length n and 1 for all entries. If we express the eigenvalues of $\tilde{\boldsymbol{K}}$ with $\lambda_1 \geq \lambda_2 \geq \cdots \geq \lambda_n$ and the corresponding set of eigenvectors $\tilde{\boldsymbol{a}}^1, \ldots, \tilde{\boldsymbol{a}}^n$, the principal component axes can be expressed as $\tilde{\boldsymbol{v}}^k = \sum_{i=1}^n \tilde{a}_i^k \phi(\boldsymbol{x}_i)$ with $\tilde{\boldsymbol{v}}^k$ and $\tilde{\boldsymbol{a}}^k$ orthonormal in $\mathcal{H}$, $k = 1, \ldots, q$ and $q$ the number of retained components. Thus solving $n \lambda \tilde{\boldsymbol{a}} = \tilde{\boldsymbol{K}} \tilde{\boldsymbol{a}}$, it is possible to compute the projection of the points into the subspace of the feature space spanned by the eigenvectors. The projection of a generic point $\boldsymbol{x}$ into this subspace becomes then $\rho_k := \langle \tilde{\boldsymbol{v}}^k, \phi(\boldsymbol{x}) \rangle = \sum_{i=1}^n \tilde{a}_i^k k(\boldsymbol{x}, \boldsymbol{x}_i)$.

Likewise, utilizing the concise, explicit form of the centered gram matrix $\tilde{\boldsymbol{K}}$, it is possible to express the projection of an arbitrary point $\boldsymbol{x}$ into the subspace spanned by the eigenvectors $\tilde{\boldsymbol{v}}^k$. Defining $\boldsymbol{Z} = (k(\boldsymbol{x}, \boldsymbol{x}_i))_{n \times 1}$, we can express this projection with the $1 \times q$ row vector

$$\rho_k = \left(Z^T - \frac{1}{n}\mathbf{1}_n^T K\right)\left(I_n - \frac{1}{n}\mathbf{1}_n\mathbf{1}_n^T\right)\tilde{\nu}, \tag{1}$$

with $\tilde{\nu}$ being the $n \times q$ matrix with the eigenvectors $\tilde{\nu}^1 \ldots, \tilde{\nu}^q$ as columns.

As we observe, the kernel PCA algorithm can be mathematically represented using only the entries of the kernel matrix. This means that the algorithm operates entirely on the original input data without requiring the computation of the new coordinates in the feature space. This technique effectively resolves the issue of potentially high computational complexity by allowing the input points to be implicitly mapped into the feature space.

However, it also introduces new challenges in terms of interpretation. Determining which input variables have the most significant impact on the kernel principal components can be highly challenging, making it difficult to interpret them in terms of the original features. In other words, since the kernel function maps the data to a higher-dimensional feature space, it can be hard to understand how the original features contribute to the newly obtained kernel principal components.

In the previous section, we have mentioned the few techniques available in the literature that can be used to gain insight into the original input variables that had the most influence on the KPCA solution. The following section presents our contribution to providing practitioners with a data-driven and faster variable ranking methodology.

**Improvement of KPCA interpretability with KPCA-IG**

It is known that gradient descent is one of the most common algorithms for the training phase of most neural networks [55]. The norm of the cost function gradient plays a crucial role as it contributes to the step size for each iteration, together with its direction and the learning rate.

Consequently, for explainability in computer vision classification models, gradient-based methods are a widespread approach for many networks, such as deep neural networks (DNN) and convolutional neural networks (CNNs). Some of the most used techniques are presented in the review proposed in [46], such as Saliency Maps [64], Deconvolutional Networks [76], Guided Backpropagation [67], SmoothGrad [65], Gradient-Input [3] and Integrated Gradients [68]. In post hoc explainability, they are often preferred over perturbation-based methods since they are not only less computationally expensive but should also be prioritized when a solution robust to input perturbation is required [46]. In the Deep Learning (DL) field, the starting point behind all the gradient-based methods is to assess the so-called attribution value of every input feature of the network. Formally, with a $p$ dimensional input $x = (x_1, \ldots, x_p)$ that produces the output $S(x) = (S_1(x), \ldots, S_C(x))$, with $C$ the numbers of output neurons, the final goal is to compute for a specific neuron $c$ the relevance of each input feature for the output. This contribution for the target neuron $c$ can be written as $R^c = (R_1^c, \ldots, R_p^c)$, as described in [2]. Depending on the method considered, the attributions are found by a specific algorithm. Generally, the gradient-based algorithms involve the computation of partial derivatives of the output $S_c(x)$ with respect to the input variables.

In the unsupervised field of KPCA, the idea cannot be applied directly as there is no classification involved, and no numeric output can be used to test the relevance of an input feature. However, as shown in [52, 58], every original variable can be represented in the input space with a function *f defined in* $\mathbb{R}^p$, representing the position of every sample point in the input space based on the values of the *p* variables.

Thus, we propose to compute at each sample point the norm of the partial derivative of every induced feature curve *projected into the eigenspace of the kernel Gram matrix.*

In support of this procedure, some works in the neuroimaging and earth system sciences domain have also shown that kernel derivatives may indicate the influence carried by the original variables as in [29, 51].

Consequently, the idea is that when the norm of the partial derivative for a variable is high, it means that the variable substantially affects the position of the sample points in the kernel PC axes. Conversely, when the norm of the partial derivative for a variable is small, the variable can be deemed negligible for the kernel principal axes.

To sum up, the novel idea of *KPCA-IG* is to find the most relevant features for the KPCA studying the projections of the gradients of the feature functions onto the linear subspace of the feature space induced by the kernel by computing the lengths of the gradient vectors with respect to each variable at each sample point as they represent how steep the direction given by the partial derivative of the induced curve is.

For completeness, we should also mention that the use of gradient in the kernel unsupervised learning framework can also be found in the context of Kernel canonical correlation analysis, as [69] proposed a new variant of KCCA that does not rely on the kernel matrix but where the maximization of canonical correlation is computed through the gradients of the pre-images of the projection directions.

Analytically we can describe our method *KPCA-IG* as follows. First, we can express the projection of *f* in the feature space through the implicit map $\phi$ as *h*. More specifically, *h* is defined on the subspace of $\mathcal{H}$ where the input points are mapped, i.e., on $\phi(\chi)$ assuming it is sufficiently smooth to support a Riemannian metric [60]. In [58], the authors demonstrated how the gradient of *h* can be expressed as a vector field in $\phi(\chi)$ under the coordinates $\boldsymbol{x} = (x^1, \ldots, x^p)$ as

$$\mathrm{grad}\left(h\right)^j = \sum_{b=1}^{p} g^{jb}(\boldsymbol{x}) D_b f(\boldsymbol{x}), \tag{2}$$

where $j = 1, \ldots, p$, $D_b$ is the partial derivative with respect to the *b* variable and $g^{jb}$ is the inverse of the Riemannian metric induced by $\phi(\chi)$ i.e. the symmetric metric tensor $g_{jb}$ which is unknown and it can be written solely in terms of the kernel [61].

The idea is to look for the curves *u* whose tangent vectors in *t* are $u'(t) = grad(h)$ as they give an indication of the local maximum variation directions of *h*.

In the previous section, we showed how to represent the projection of every mapped generic point $\phi(\boldsymbol{x})$ into the subspace spanned by the eigenvectors of $\tilde{\boldsymbol{K}}$ in (1).

Similarly, the $u(t)$ curves can be projected into the subspace of the kernel PCA. We define $u(t) = k(\cdot, \boldsymbol{x}(t))$ with $\boldsymbol{x}(t)$ the solution of $\frac{dx^j}{dt} = \mathrm{grad}\left(\mathrm{h}\right)^j$ and $\boldsymbol{Z}_t = (k(\boldsymbol{x}(t), \boldsymbol{x}_i))_{n \times 1}$
.

Now we can define the induced curve in the KPCA axes with the row vector:

$$\varphi_{1\times q} = \left(\mathbf{Z}_t^T - \frac{1}{n}\mathbf{1}_n^T \mathbf{K}\right)\left(\mathbf{I}_n - \frac{1}{n}\mathbf{1}_n\mathbf{1}_n^T\right)\tilde{\mathbf{v}}. \tag{3}$$

In order to assess the influence of the original variables on the coordinates of the data points into the kernel principal axes, we can represent the gradient vector field of $h$ i.e. the tangent vector field of $u(t)$ into the KPCA solution. Formally, the tangent vector at an initial condition $t = t_0$ with $x_0 = \phi^{-1} \circ u(t_0)$ can be obtained as $\frac{du}{dt}|_{t=t_0}$ and the projected directions of maximum variation as

$$\begin{aligned}w_{1\times q} &= \frac{d\varphi}{dt}\bigg|_{t=t_0} = \frac{d\mathbf{Z}_t^T}{dt}\bigg|_{t=t_0}\left(\mathbf{I}_n - \frac{1}{n}\mathbf{1}_n\mathbf{1}_n^T\right)\tilde{\mathbf{v}}\\ &= \left[\frac{d\mathbf{Z}_t^1}{dt}\bigg|_{t=t_0},\dots,\frac{d\mathbf{Z}_t^n}{dt}\bigg|_{t=t_0}\right]^T\left(\mathbf{I}_n - \frac{1}{n}\mathbf{1}_n\mathbf{1}_n^T\right)\tilde{\mathbf{v}},\end{aligned} \tag{4}$$

with

$$\begin{aligned}\frac{d\mathbf{Z}_t^i}{dt}\bigg|_{t=t_0} &= \frac{dk(\mathbf{x}(t),\mathbf{x}_i)}{dt}\bigg|_{t=t_0}\\ &= \sum_{j=1}^p D_j k(\mathbf{x}_0,\mathbf{x}_i)\frac{dx^j}{dt}\bigg|_{t=t_0}.\end{aligned} \tag{5}$$

If we assume that $\phi(\chi)$ is flat (Euclidean subspace), the metric tensor $g_{jb}$ becomes the Kronecker delta $\delta_{jb}$ which is equal to 0 for $j \neq b$ and to 1 when $j = b$.

In this case, (2) becomes

$$\text{grad}\left(h\right)^j = D_j f(\mathbf{x}) \tag{6}$$

and (5):

$$\begin{aligned}\frac{d\mathbf{Z}_t^i}{dt}\bigg|_{t=t_0} &= \frac{dk(\mathbf{x}(t),\mathbf{x}_i)}{dt}\bigg|_{t=t_0}\\ &= D_j k(\mathbf{x}_0,\mathbf{x}_i) D_j f(\mathbf{x}).\end{aligned} \tag{7}$$

If we further assume that $f$ takes the linear form $f(\mathbf{x}) = \mathbf{x} + t\mathbf{e}_j$ with $t \in \mathbb{R}$ and $\mathbf{e}_j = (0,\dots,1,\dots,0)$, having the value 1 only for the $j$-th component, we obtain the same expression as in [52]

$$\begin{aligned}\frac{d\mathbf{Z}_t^i}{dt}\bigg|_{t=0} &= \frac{dk(\mathbf{x}(t),\mathbf{x}_i)}{dt}\bigg|_{t=0}\\ &= D_j k(\mathbf{x},\mathbf{x}_i).\end{aligned} \tag{8}$$

Thus, with these hypotheses we can use (4) to compute the projected directions of maximum variation with respect to the $j$-th variable as

$$w^j_{1\times q} = \frac{d\varphi^j}{dt}\bigg|_{t=0} = \frac{d\mathbf{Z}_t^T}{dt}\bigg|_{t=0}\left(\mathbf{I}_n - \frac{1}{n}\mathbf{1}_n\mathbf{1}_n^T\right)\tilde{\mathbf{v}}, \tag{9}$$

with $\left.\frac{d\mathbf{Z}_t^i}{dt}\right|_{t=0}$ as in (8).

If we take as kernel the radial basis kernel $k(\mathbf{x}, \mathbf{x}_i) = exp(-\sigma \|\mathbf{x} - \mathbf{x}_i\|^2)$, then (8) becomes, as showed in [52]:

$$\left.\frac{d\mathbf{Z}_t^i}{dt}\right|_{t=0} = -2\sigma k(\mathbf{x}, \mathbf{x}_i)(x^j - x_i^j), \tag{10}$$

with $i = 1, \ldots, n$ and with $x^j$ the value for the variable $j$ for the generic point $\mathbf{x}$.

If we consider that the $1 \times q$ row vector $w^j$ can be computed for all the training points rather than only for a generic point $\mathbf{x}$, we obtain a $n \times q$ matrix $\mathbf{W}^j$ giving the direction of maximum variation associated with the $j$-th variable for each input point.

Thus, the idea is to first compute the norm of this partial derivative with respect to the variable $j$ for each sample point and then compute the mean value of these $n$ contributions. The score that we obtain suggests the relevance of the $j$ variable in the KPCA solution.

Analytically, the $n \times q$ matrix $\mathbf{W}^j$ rows are denoted by the $1 \times q$ vectors $\mathbf{v}_i^j$ as $\mathbf{v}_i^j = (w_{i1}^j, \ldots, w_{iq}^j)$ where $w_{ik}^j$ denotes the entry in the $i$-th row and $k$-th column of $\mathbf{W}^j$. The square root of the quadratic norm of $\mathbf{v}_i^j$ is given by

$$\|\mathbf{v}_i^j\| = \sqrt{\|\mathbf{v}_i^j\|^2} = \sqrt{\sum_{k=1}^q \left(w_{ik}^j\right)^2}. \tag{11}$$

We can now compute the square root of the quadratic norms for all $n$ rows, resulting in $n$ values $\|\mathbf{v}_1^j\|, \ldots, \|\mathbf{v}_n^j\|$. The mean of these values is given by

$$r^j = \frac{1}{n} \sum_{i=1}^n \|\mathbf{v}_i^j\| = \frac{1}{n} \sum_{i=1}^n \sqrt{\sum_{k=1}^q \left(w_{ik}^j\right)^2}. \tag{12}$$

Thus, $r^j$, the mean of the norm vectors of the partial derivative of $\varphi^j$ among all the $n$ sample points, it can give an indication of the overall influence of the $j$-th variable on the points.

Finally, we can repeat the procedure for all the p variables with $j = 1, \ldots, p$. The vector $\mathbf{r} = (r^1, \ldots, r^p)$ will contain all the mean values of the norm vectors for every of the $p$ variables, and after sorting them in descending order, it represents the ranking of the original features proposed by KPCA-IG. Every entry of $\mathbf{r}$ is a score that indicates the impact of every variable on the kernel PCA representation of the data, from the most influential to the least important.

The method is non-iterative, and it only requires linear algebra. Thus, it is not susceptible to numerical instability or local minimum problems. It variables representing the ranking of the original features proposed by KPCA-IG. Every entry of*rr*is a score that indicates the impact of every variable on the kernel PCA representation of the data, it is then sorted decreasingly, from the most influential to the least important.is computationally very fast, and it can be applied to any kernel function that admits a first-order derivative. The described procedure has been implemented on R, and the code can be available upon request to the authors.

## Results

We conducted experiments on three benchmark datasets from the biological domain to assess the accuracy of the proposed unsupervised approach for feature selection. These datasets include two microarray datasets, named *Carcinom* and *Glioma*, which are available in the Python package scikit-feature [35] and the gene expression data from normal and prostate tumour tissues [10], *GPL93* from the GEO, a public functional genomics data repository. *Glioma* contains the expression of 4434 genes for 50 patients, while *Carcinom* 9182 genes for 174 individuals. Both datasets have already been used as a benchmark in numerous studies including several methods comparisons, such as [7, 35]. Then, the dataset *GPL93* contains the expression of 12626 genes for 165 patients, and it has been chosen for its complexity and higher dimensionality.

The idea is to compare the proposed methodology *KPCA-IG* with existing unsupervised feature selection methods from diverse frameworks, as conducted in [7, 35]:

- *lapl* [23], to include one method that relies on the computation of a score.
- *NDFS* [37], to add one of the methods primarily designed for clustering. It is based on the implicit assumption that samples are structured into subgroups and demands the a priori definition of the number of clusters.
- *KPCA-permute* in [40] available in the *mixKernel* R package to include another methodology from the context of kernel PCA.

To evaluate the selected features provided by the four methods, we measured the overall accuracy (ACC) and normalized mutual information (NMI) [15] based on k-means cluster performance. For each method, the k-means clustering ACC and NMI have been obtained using several subsets with a different number $d$ of selected features, with $d \in \{10, 20, \ldots, 290, 300\}$. Thus, the relevance of the selected values has been estimated according to their ability to reconstruct the clustered nature of the data. More specifically, the three datasets *Glioma*, *Carcinom* and *GPL93* are characterized by 4, 11 and 4 groups respectively. Thus, the k-means clustering was computed using the correct number of clusters in the datasets to obtain a metric for the capability of the selected features to keep this nature.

Note that only the *NDFS* method is implemented to explicitly obtain an optimum solution in terms of clustering, also requiring in advance the number of groups in the data. For each method, the k-means clustering was run 20 times to obtain a mean of the overall accuracy and normalized mutual information for each of the 30 subsets of selected features. Both our novel method *KPCA-IG* and *KPCA-permute* have been employed with a Gaussian kernel with a sigma value depending on the dataset. The selected features are, in both cases, based on the first 3, 5 and 3 kernel PC axes for *Glioma*, *Carcinom* and *GPL93*, respectively. The CPU time in seconds required to obtain the feature ranking for all the methods has also been observed. The experiment was conducted on a standard laptop Intel Core *i*5 with 16*GB* RAM.

### Evaluation on benchmarks datasets

In Table 1, we can see the results in terms of mean Accuracy and NMI over 20 runs for different numbers of retained features $d$. For the first dataset *Glioma lapl* seems

Table 1 Comparison of the different methods in terms of mean ACC and NMI over 20 runs of a k-means clustering for several subsets with a different number d of selected features

|  | lapl | NDFS | KPCA-permute | KPCA-IG |
|---|---|---|---|---|
| *Glioma* ($n = 50, p = 4434$) | | | | |
| ACC(10) | **0.50** (0.02) | 0.37 (0.04) | 0.48 (0.03) | 0.42 (0.01) |
| NMI(10) | **0.34** (0.02) | 0.13 (0.03) | 0.31 (0.02) | 0.21 (0.01) |
| ACC(150) | **0.56** (0.03) | 0.53 (0.04) | 0.54 (0.04) | **0.56** (0.05) |
| NMI(150) | **0.50** (0.02) | 0.41 (0.03) | 0.48 (0.02) | 0.36 (0.05) |
| ACC(300) | 0.54 (0.04) | 0.55 (0.04) | 0.52 (0.03) | **0.57** (0.05) |
| NMI(300) | **0.48** (0.03) | 0.41 (0.03) | 0.45 (0.02) | 0.35 (0.05) |
| CPU time | 0.4 | 84.6 | 620.9 | 2.9 |
| *Carcinom* ($n = 174, p = 9182$) | | | | |
| ACC(10) | 0.27 (0.02) | 0.47 (0.04) | 0.48 (0.02) | **0.51** (0.01) |
| NMI(10) | 0.23 (0.01) | 0.48 (0.03) | 0.43 (0.01) | **0.49** (0.02) |
| ACC(150) | 0.61 (0.03) | 0.68 (0.04) | 0.67 (0.03) | **0.70** (0.02) |
| NMI(150) | 0.62 (0.03) | **0.72** (0.03) | 0.69 (0.03) | 0.70 (0.03) |
| ACC(300) | 0.69 (0.04) | 0.69 (0.04) | **0.70** (0.048) | 0.69 (0.03) |
| NMI(300) | 0.73 (0.03) | **0.73** (0.03) | 0.71 (0.03) | 0.70 (0.02) |
| CPU time | 1.4 | 391.8 | 7937.6 | 30.5 |
| *GPL93* ($n = 165, p = 12626$) | | | | |
| ACC(10) | 0.38 (0.01) | 0.41 (0.01) | **0.42** (0.01) | 0.40 (0.01) |
| NMI(10) | 0.08 (0.01) | 0.109 (0.07) | 0.11 (0.01) | **0.15** (0.01) |
| ACC(150) | 0.38 (0.01) | 0.45 (0.01) | **0.60** (0.02) | 0.58 (0.01) |
| NMI(150) | 0.07 (0.01) | 0.18 (0.02) | **0.39** (0.01) | 0.29 (0.01) |
| ACC(300) | 0.37 (0.01) | 0.49 (0.03) | 0.56 (0.05) | **0.56** (0.01) |
| NMI(300) | 0.07 (0.01) | 0.22 (0.02) | 0.22 (0.01) | **0.31** (0.01) |
| CPU time | 2.1 | 1277.4 | 17691.4 | 39.8 |

CPU represents the computational time in seconds required by the four methods only to find the most influential features to show the best performance in terms of NMI and AUC, except when $d = 300$ where the Accuracy obtained with *KPCA-IG* is the highest, even if all the methods seem to behave very similarly in terms of ACC. Analyzing the results for the other two datasets *Carcinom* and *GPL93* that are considerably bigger and possibly more complex in terms of sample space manifold, the two methods based on the kernel framework exhibit to surpass the *lapl* and *NDFS* approaches, especially in the *GPL93* datasets. The comparison of the different approaches in terms of NMI and ACC of these two datasets can also be observed in Figs. 1 and 2.

Moreover, as shown in [7] *NDFS* and the other cluster-based methods like *MCFS* and *UDFS* suffer if the user selects an incorrect decision for the a priori number of clusters. In our case, we show that the proposed methodology behaves similarly or even better to a method like *NDFS* that is specifically optimized for this cluster setting.

The two kernel-based approaches, namely *KPCA-permute* and our novel method *KPCA-IG*, reveal an excellent performance in this setting, once again displaying the appropriateness of the kernel framework in the context of complex biological datasets. However, *KPCA-IG* can provide these above-average performances with a considerably lower CPU time.

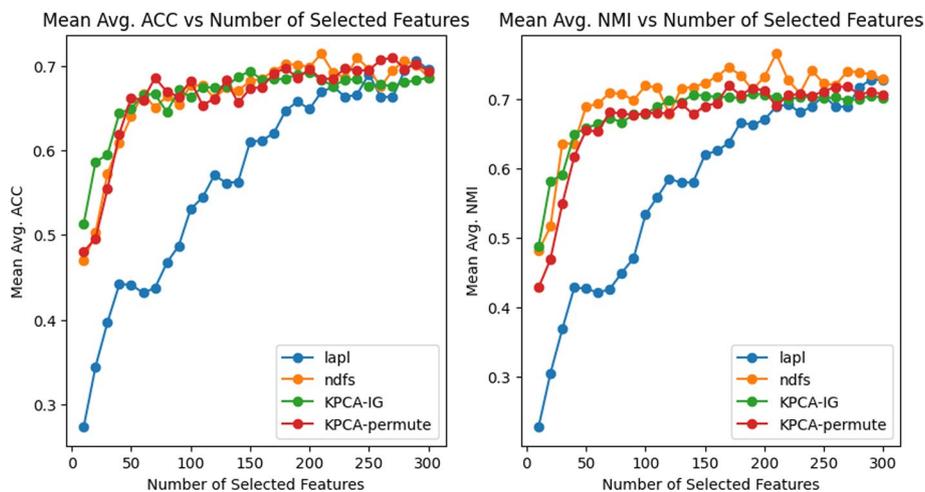

**Fig. 1** *Carcinom* ACC and NMI: Comparison of the performance of the four methods on the *Carcinom* dataset in terms of Accuracy (left) and Normalized Mutual Information (right) as a function of the number of selected features *d*. ACC and NMI are computed for the k-means results using only the d selected features

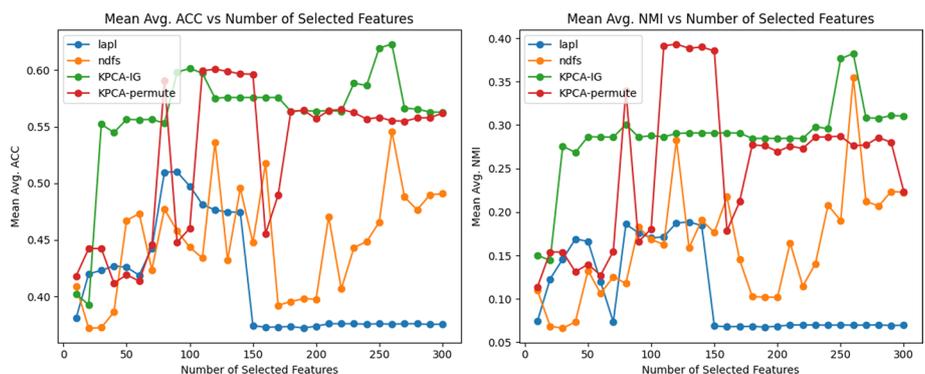

**Fig. 2** *GPL93* ACC and NMI: Comparison of the performance of the four methods on the *GPL93* dataset in terms of Accuracy (left) and Normalized Mutual Information (right) as a function of the number of selected features *d*. ACC and NMI are computed for the k-means results using only the d selected features

Only *lapl* seems as fast as *KPCA-IG* while showing poorer results in the more complex scenario represented in this case by the *GPL93* dataset. Other methods, such as the concrete autoencoder in [1], have proven successful in this context. The results obtained with the concrete autoencoder, as demonstrated in [7], were comparable or even inferior in terms of accuracy and NMI. Furthermore, the computational time required to achieve these results was on the order of days. As a result, we opted not to include it in our simulations.

**Application on Hepatocellular carcinoma dataset**

Liver cancer is a global health challenge, and it is estimated that there will be over 1 million cases by 2025. Hepatocellular carcinoma (HCC) is the most common type of liver cancer, accounting for around 90% of cases [39].

The most significant risk factors associated with HCC are, among others, chronic hepatitis B and C infections, nonalcoholic fatty disease, and chronic alcohol abuse [44].

To analyse the use of the *KPCA-IG* method, we used the expression profiling by array of an HCC dataset from the Gene expression Omnibus (series GSE102079). It contains the gene expression microarray profiles of 3 groups of patients. First, 152 patients with HCC who were treated with hepatic resection between 2006 and 2011 at Tokyo Medical and Dental University Hospital. Then, the gene expression of normal liver tissues of 14 patients as control [12]. The third group contains the gene expression of 91 patients with liver cancer but of non-tumorous liver tissue. The total expression matrix for the 257 patients contains the expression of 54613 genes, and the data has been normalised by robust multichip analysis (RMA) as in [19] and scaled and centered before applying KPCA.

To show the potentiality of KPCA-IG, we first perform kernel PCA with radial basis kernel with $\sigma = 0.00001$, which was set heuristically to maximize the explained variance and obtain a clear two dimension data representation. Even if detecting groups is not the optimization criterion of kernel PCA, it is possible to see that the algorithm catches the dataset's clustered structure in Figs. 7 and 8.

For this reason, applying a method like the proposed KPCA-IG can enlighten the kernel component axes, possibly giving an interpretation of the genes' influence on the sample points representation.

The KPCA-IG provides a feature ranking based on the KPCA solution, in this case, based on the first two kernel Principal Components. As mentioned before, one of the main advantages of the proposed method is the fast computational time required, as with this high-dimensional dataset, the CPU time was 654.1 seconds. Table 2 presents the first 25 genes and Fig. 3 the distribution of the 54613 variables scores.

To obtain an indication of the significance of the variables selected by KPCA-IG in terms of retained information, we computed the Silhouette Coefficient, a metric used to assess the goodness of a clustering solution. The score values are in $[-1, 1]$ where values close to 1 mean that the clusters are clearly separated [6, 54]. In this case, even if KPCA is not optimized to create clusters of data, a 2D KPCA plot with clear separation among different groups may suggest a solution with more explained variability.

Starting from the feature ranking given by KPCA-IG, we computed the silhouette scores for the KPCA solution for 5462 subsets in terms of original variables, i.e. datasets with the increasing number of features of $5, 15, 25, \ldots, 54605, 54613$. The metric was first computed using the features ranked by KPCA-IG and then with 5 different random feature rankings.

From Fig. 4, it can be seen that the scores obtained for kernel PC solutions applied to datasets composed by the features selected by KPCA-IG are consistently higher when compared to those where the ranking of the variables is chosen randomly.

To see more in detail the behaviour of the curves for reduced datasets with a small number of features, which can be more relevant for practical biological use, Fig. 5 represents the scores for 163 subsets of the form in terms of selected features of $5, 10, \ldots, 100, 110, \ldots, 1000, 2000, \ldots, 54000, 54613$. As we can see, the information retained by the KPCA-IG selected features is higher, as they lead to Silhouette scores for the KPCA plots closer to 1. All the obtained coefficients are based on a two cluster

**Table 2** The 25 most relevant genes and the last 3 out of the total number of 54613 according to the proposed KPCA-IG method

| Genes | Score | Standard deviation | Symbol |
|---|---|---|---|
| 1555797_a_at | 0.427972 | 0.12993 | ARPC5 |
| 237350_at | 0.426140 | 0.11331 | TTC36 |
| 1559573_at | 0.424048 | 0.11844 | LINC01093 |
| 230478_at | 0.420690 | 0.12039 | OIT3 |
| 203213_at | 0.417682 | 0.11453 | CDK1 |
| 205019_s_at | 0.417597 | 0.11383 | VIPR1 |
| 1559065_a_at | 0.417234 | 0.12463 | CLEC4G |
| 205984_at | 0.417234 | 0.12607 | CRHBP |
| 220114_s_at | 0.416410 | 0.12014 | STAB2 |
| 202604_x_at | 0.416228 | 0.12273 | ADAM10 |
| 220496_at | 0.415608 | 0.12558 | CLEC1B |
| 205866_at | 0.414893 | 0.11862 | FCN3 |
| 214895_s_at | 0.414887 | 0.13144 | ADAM10 |
| 240963_x_at | 0.413698 | 0.12648 | PLXDC1 |
| 234304_s_at | 0.413574 | 0.13119 | IPO11 |
| 222077_s_at | 0.412939 | 0.11637 | RACGAP1 |
| 223341_s_at | 0.411044 | 0.13771 | SCOC |
| 214710_s_at | 0.410616 | 0.11262 | CCNB1 |
| 218009_s_at | 0.410610 | 0.11377 | PRC1 |
| 219918_s_at | 0.410460 | 0.11532 | ASPM |
| 226524_at | 0.410119 | 0.13299 | C3orf38 |
| 201890_at | 0.410097 | 0.11593 | RRM2 |
| 207804_s_at | 0.409962 | 0.12230 | FCN2 |
| 210481_s_at | 0.409839 | 0.12106 | CLEC4M |
| 209470_s_at | 0.409759 | 0.12423 | GPM6A |
| ... | ... | | |
| 229461_x_at | 0.0520878 | 0.088739 | NEGR1 |
| 230538_at | 0.0520119 | 0.134812 | SHC4 |
| 206145_at | 0.0507935 | 0.098227 | RHAG |

The original scores and their standard deviations have been multiplied by $10^3$ for a better visualization solution where the sigma parameter of the Gaussian kernel has been adapted to the different numbers of considered features to obtain a KPCA solution with maximized explained variance.

Moreover, to check the generalization properties of KPCA-IG, we compared the explained variance captured by the selected variables on training and test data. More specifically, based on 5 different random train-test splits (192 train and 65 test points), we computed the feature ranking using KPCA-IG for each of the 5 training sets. Based on these features, we plotted the variance explained by KPCA. Again the analysis is made on different subsets with an increasing number of variables retained, namely $5, 10, \ldots, 100, 110, \ldots, 1000, 2000, \ldots, 54000, 54613$.

Now, the variable ranking obtained by KPCA-IG on the training set is applied on the test sets. Thus, the computed KPCA explained variance on the test sets containing only those features is compared to the one obtained with KPCA applied on the training sets. In Fig. 6, it can be seen that the variance explained by KPCA for all the subsets of selected features on the training and the test sets is consistently similar for all of the 5 randomly generated train-test splits.

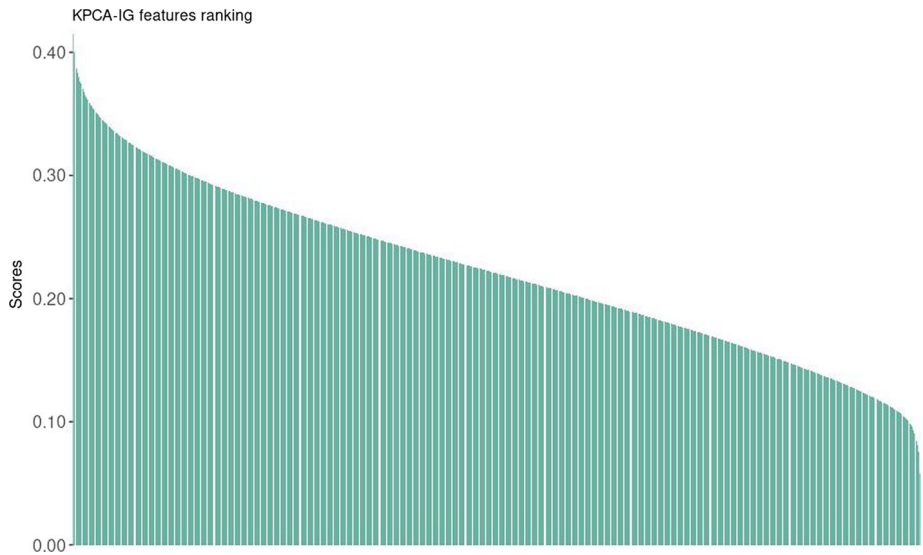

**Fig. 3** Distribution of the scores for the ordered 54613 genes, from a maximum of 0.428 to a minimum of $0.05 \times 10^{-3}$

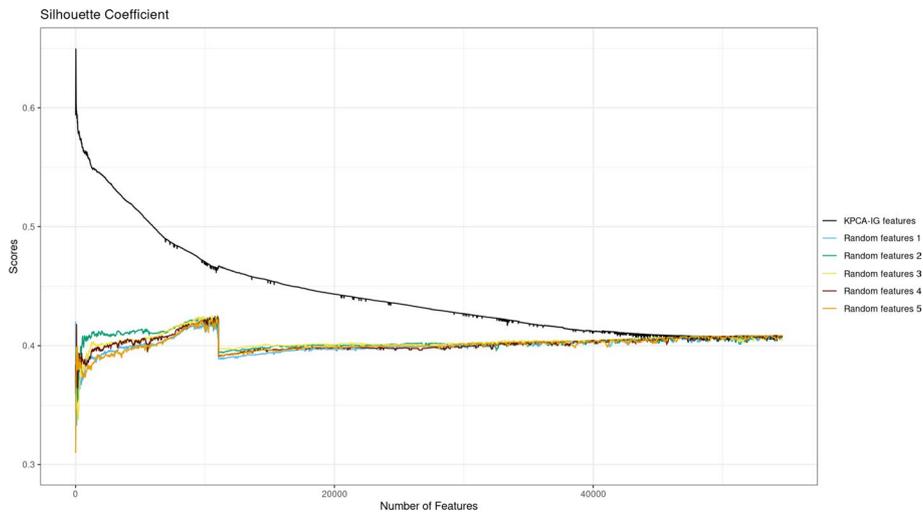

**Fig. 4** Silhouette Coefficients for the KPCA applied on datasets with features ranked according to KPCA-IG vs 5 different randomly generated rankings

Another way to assess the relevance of the obtained ranking is to visualize the genes with the method proposed by [52] and to see if the bio-medical community has already investigated the retained genes.

For instance, Fig. 7 displays the representation of the variable TTC36, the second feature in the ranking provided by KPCA-IG as it is the first gene that shows differential expression.

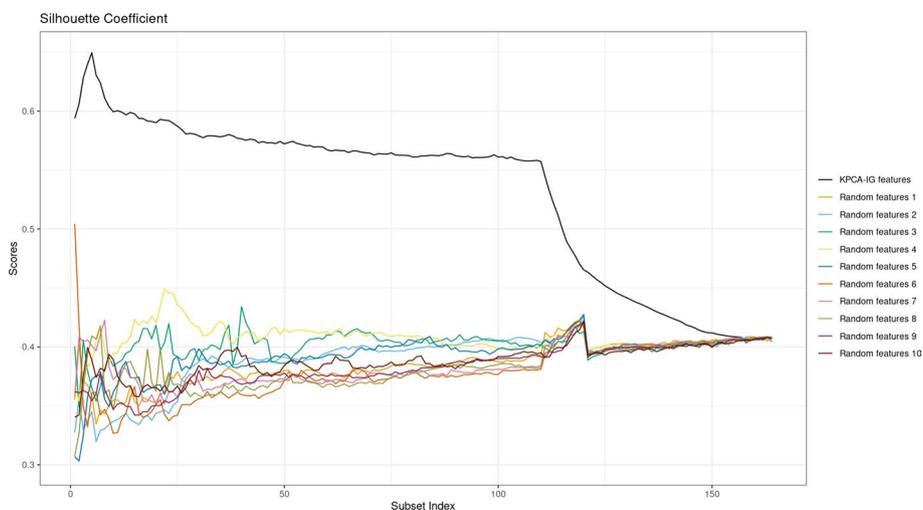

**Fig. 5** Silhouette Coefficients for the KPCA applied on datasets with features ranked according to KPCA-IG vs 10 different randomly generated rankings, with particular attention given to subsets with a small number of features

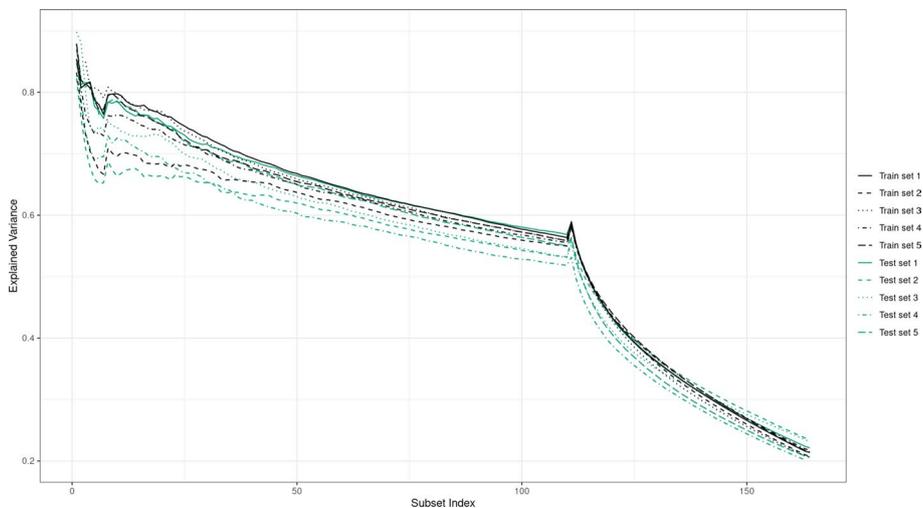

**Fig. 6** Variance explained by the KPCA based on different subsets of retained features by KPCA-IG on the training sets. Comparison between train and test sets

The direction of the arrows suggests an upper expression of the gene towards the cluster of patients that do not have liver cancer or patients whose liver tissue is not tumorous.

To validate the procedure, we selected relevant literature about the gene TTC36. This gene, also known as HBP21, is a protein encoding gene. It has been shown that this gene's encoded protein may function as a tumour suppressor in hepatocellular carcinoma (HCC) since it promotes apoptosis while it has been proven to be downregulated in HCC cases [27].

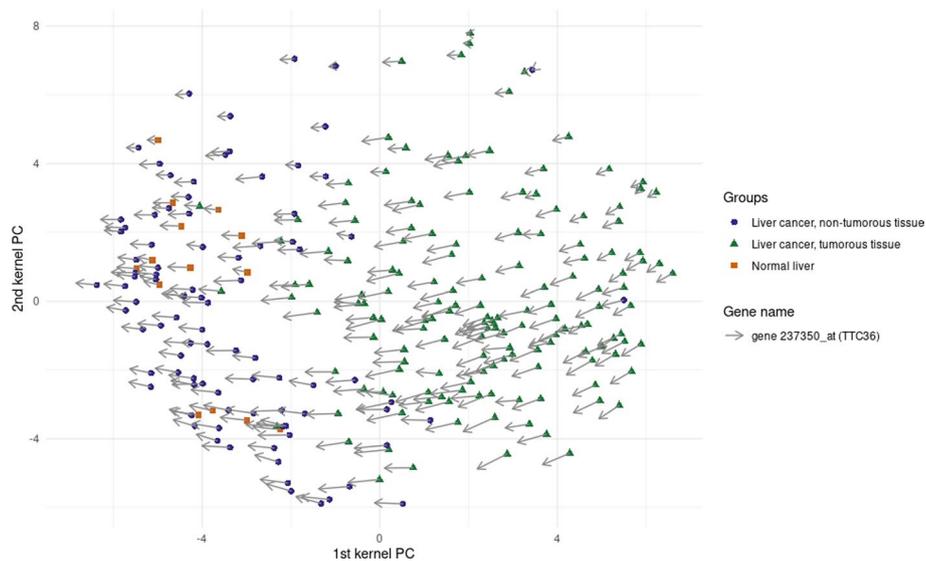

**Fig. 7** Gene TTC36: Gene visualization obtained using the procedure described in [52]

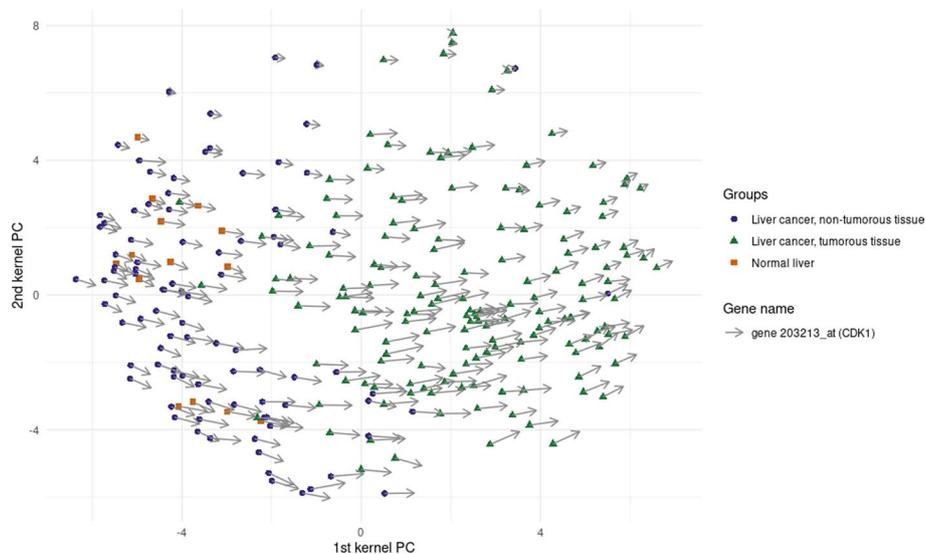

**Fig. 8** Gene CDK1: Gene visualization obtained using the procedure described in [52]

Another gene that shows differential expression in the two groups is CDK1. In this case, Fig. 8 suggests that this gene seems to be upregulated in the presence of cancer tissue.

The indication found in multiple studies is that the increased expression of this gene is indeed linked with a poorer prognosis or outcome, such as high tumour grade, invasion of lymphovascular or muscularis propria, and the presence of distant metastasis [24, 34, 36, 38, 66]. In the same way, another of the most critical genes, according to KPCA-IG, that seems to be prominent in the case of an HCC patient reflecting the same indication in the medical literature is ADAM10, known to be involved in the RIPing and shedding of numerous substrates leading to cancer progression and inflammatory disease [31], and indicated as a target for cancer therapy [13, 45], while being upregulated in metastasis cancers [20, 33]. Rac GTPase activating protein 1 gene RACGAP1 selected by KPCA-IG shares a similar behaviour with ADAM10 and CDK1. The literature concerning this gene is also broad, where it has been marked as a potential prognostic and immunological biomarker in different types of cancer, such as gastric cancer [56], uterine carcinosarcoma [42], breast cancer [49] or colorectal cancer [26] among many others. CCNB1 has also been indicated to be an oncogenic factor in the proliferation of HCC cells [9], showing a significant impact on the patient's survival time [16, 79] and thus has been targeted for cancer treatments [18]. PRC1 has revealed upper expression in other cancer tissues such as, among others, invasive cervical carcinomas [57], papillary renal cell carcinoma [74], pediatric adrenocortical tumour [72], while yet not being studied in depth as compared to CCNB1, ADAM10 or RACGAP1.

ASPM was known initially as a gene involved in the control of the human brain development and in the cerebral cortical size [5, 77] whose mutations may lead to primary autosomal recessive microcephaly [30], more recently its overexpression has also been linked with tumour progression as in [71, 73].

Lastly, for the group of upregulated genes in HCC, RRM2 has also been linked with low overall survival [11, 28], leading to exhaustive cancer research suggesting targeting its inhibition for different types of tumour treatments [47, 48, 50, 70].

On the other hand, the selected genes that manifest down-regulation in cancerous HCC tissues are LINC01093, OIT3, VIPR1, CLEC4G, CRHBP, STAB2, CLEC1B, FCN3, FCN2 and CLEC4M. The literature regarding these genes indicates that they work as suppressors in different cancerous situations, once again endorsing the selection provided by KPCA-IG for the upregulated genes.

The few genes in the first 25 selected by KPCA-IG that do not exhibit differential expression using [52] method (ARPC5, IPO11, C3orf38, SCOC) are potential genes that explain much variability in the data or that share a possibly nonlinear interaction with the differential expressed genes. Since the ultimate goal of KPCA is not to discriminate groups, it is expected that some of the variables found by the novel method are not linked with a classification benefit. However, further follow-up on the function of these genes may be done in cooperation with an expert in the field.

## Conclusion

We have seen how the unsupervised feature selection literature is narrower than its supervised counterpart. Moreover, algorithms that use the kernel principal component analysis for feature selection are reduced to a few works. In the present work, we have introduced a novel method to enhance variables' interpretability in kernel PCA. Using benchmark datasets, we have proven the comparability in terms of accuracy with already existing and recognized methods, where the efficiency of KPCA-IG has proven to be competitive. The application on the real-life Hepatocellular carcinoma dataset and the validation obtained from the comparison of the selected variables by the method with the bio-medical literature have confirmed the effectiveness and strengths of the proposed methodology. In future works, further in-depth analysis will be realized to assess the impact of the choice of the kernel function on the feature ranking obtained by KPCA-IG. Moreover, the method will be adapted to other linear algorithms that are solely based on dot-products hence supporting a kernelized version, such as kernel Discriminant Analysis or kernel Partial Least-Squares Discriminant Analysis.

#### Abbreviations
| | |
|---|---|
| KPCA | Kernel principal component analysis |
| KPCA-IG | Kernel principal component analysis Interpretable Gradient |
| HCC | Hepatocellular carcinoma |
| SPEC | Spectral Feature Selection |
| MCFS | Multi-Cluster Feature Selection |
| NDFS | Nonnegative Discriminative Feature Selection |
| UDFS | Unsupervised Discriminative Feature Selection |
| CPFS | Convex Principal Feature Selection |
| lapl | Laplacian score |

#### Acknowledgements
The authors thank Dr. Alberto González-Sanz for the help with the revision of the mathematical formulations and Gabriele Tazza for the valuable and enriching discussions.

#### Author contributions
MB wrote the main manuscript and performed the analysis under the supervision of SD and MAD. All authors read and approved the final manuscript.

#### Funding
This work was funded by e-MUSE MSCA-ITN-2020 European Training Network under the Marie Sklodowska-Curie Grant Agreement No. 956126. The funding did not influence the study's design, the interpretation of data, or the writing of the manuscript.

#### Availability of data and materials
Glioma and Carcinoma datasets are freely available at https://github.com/jundongl/scikit-feature/tree/master/skfeature/data, GPL93 is freely available at https://www.ncbi.nlm.nih.gov/geo/query/acc.cgi?acc=GPL93 HCC datatset is freely available at https://www.ncbi.nlm.nih.gov/geo/query/acc.cgi?acc=GSE102079.

#### Declarations

#### Ethics approval and consent to participate
Not applicable.

#### Consent for publication
Not applicable.

#### Competing interests
The authors declare that they have no competing interests.

Received: 27 March 2023   Accepted: 27 June 2023

Published online: 12 July 2023

## Publisher's Note

Springer Nature remains neutral with regard to jurisdictional claims in published maps and institutional affiliations.